\documentclass[11pt,a4paper]{article}
\usepackage{amsfonts}

\newcommand{\vsp}{$\vphantom{\Big |}$}
\newcommand{\RR}{\mathbb{R}}
\newcommand{\CC}{\mathbb{C}}
\newcommand{\ZZ}{\mathbb{Z}}

\newcommand{\QQ}{\mathbb{Q}}
\newcommand{\hf}{{\textstyle\frac{1}{2}}}
\newcommand{\bw}{{\textstyle\bigwedge}}
\newcommand{\Wp}{{\textstyle\bigwedge^+}}
\newcommand{\Wm}{{\textstyle\bigwedge^-}}
\newcommand{\DD}{I\kern-3.5pt D}
\newcommand{\FF}{I\kern-3.5pt F}
\newcommand{\arr}[1]{\smash{\mathop{\longrightarrow}\limits^{#1}}}
\begin{document}
\begin{center}
\LARGE\bf       Superconnections:\\ 
      an Interpretation of the Standard Model\footnote{Talk presented at
      the Wichmann Symposium held at UC Berkeley, June 1999}
\end{center}
\vspace{3mm}
\begin{center}\large 
       Gert Roepstorff\\
       Institute for Theoretical Physics\\
       RWTH Aachen\\
       D-52062 Aachen, Germany\\
       e-mail: roep@physik.rwth-aachen.de\\[7mm]\normalsize
       Dedicated to Eyvind H.\ Wichmann\\
       on occasion of his 70th birthday
\end{center}
\vspace{5mm}\par\noindent
{\bf Abstract}. The mathematical framework of superbundels as
pioneered by D.\ Quillen suggests that one considers the Higgs field
as a natural constituent of a superconnection. I propose to take as
superbundle the exterior algebra obtained from a Hermitian vector bundle
of rank $n$ where $n=2$ for the electroweak theory and $n=5$ for the
full Standard Model. The present setup is similar to but avoids the use
of non-commutative geometry.

\section{Introduction}
The key to our present-day understanding of the electroweak interactions
is the spontaneous breakdown of local gauge symmetries. However, the mass 
generating mechanism requires the introduction of the so-called Higgs field.
A long-standing problem is to give meaning to scalar fields as natural
ingredients of a gauge theory. The subject has received special attention,
since, up to now, the Higgs particle has not been observed in experiments.
It would be impossible to provide a coherent account of all attempts to
interpret the Higgs field within the context of supersymmetry or
non-commutative geometry, nor shall I try to review the history of the
Standard Model, or discuss its details. In the present approach, which I 
believe is new, I continue the work begun in [2,3] and concentrate on one 
aspect only: the possible use of Quillen's concept of a superconnection [1] 
in physics, since it became increasingly clear to me that Euclidean field 
theory is the study of $G$ superbundles.
The goals that motivate such a study are:
\begin{itemize}
\item To reduce the number of free parameters of the Standard Model
\item To think of the Higgs field as some extension of the conventional
gauge potential
\item To naturally explain the form of the Higgs potential
\item To unite the gauge coupling and the Yukawa coupling to fermions
in one Lagrangian, $\bar{\psi}i{\cal D}\psi$, where $\cal D$ is a generalized 
Dirac operator
\item To predict the mass of the Higgs boson
\item To predict the number of fermion generations and the structure of 
      the Cabibbo-Kobayashi-Maskawa (CKM) matrix.
\end{itemize}
Let us start with a few definitions. By a {\em superspace\/} we mean
a $\ZZ_2$-graded vector space $V=V^+\oplus V^-$. Elements of $V^\pm$ are 
said to be
\begin{quote}
          --  even/odd,\\ 
          --  right-handed/left-handed,\\
          --  positive/negative,\\
          --  matter-/antimatter-like, or\\ 
          --  bosonic/fermionic
\end{quote}
depending on their use in physics. Examples of spaces with such a structure
are abundant in the theory of elementary particles. In most instances,
${\rm dim\,}V^+={\rm dim\,}V^-$. For brevity we shall refer to the
even(odd)ness indicated by the $\pm$ sign as the {\em parity\/} of elements
in $V$. Notice also that the $\ZZ_2$-grading carries
over to direct sums and tensor products of graded vector spaces in an
obvious manner. 

A {\em superalgebra\/} is a superspace whose product respects the grading, 
i.e.\ the even(odd)ness of its elements. Examples are:
\begin{quote}-- the exterior algebra of an ungraded vector space,\\
             -- the Clifford algebra of an ungraded vector space,\\
             -- the endomorphism algebra of a superspace.
\end{quote}
Exterior algebras will be seen to play a particular role in what follows.
We therefore remind the reader that the exterior algebra $\bw E$ of a
vector space $E$ is $\ZZ$-graded by the degree $p$ of the exterior power
and $\ZZ_2$-graded by the parity $(-1)^p$. Notice that 
${\rm dim}\,\Wp E={\rm dim}\,\Wm E=2^{n-1}$ where $n={\rm dim}\,E$.

Within a superalgebra $A$ the {\em supercommutator\/} is defined as follows:  
$$ 
   [a,b] =\cases{ab+ba & if $a,b$ are odd\cr ab-ba & else\cr}
   \qquad(a,b\in A).
$$
Hence, the supercommutator of a pair of odd elements is in fact their
anticommutator. From now on brackets $[\cdot,\cdot]$ will always denote
the supercommutator provided the parity of its arguments are
unambigeously defined. By construction, any exterior algebra is
supercommutative, i.e., all brackets vanish. One calls {\em supertrace\/} 
any linear functional that vanishes on supercommutators. With exterior
algebras any linear functional is a supertrace.

When it comes to studying differential operators on manifolds, the concept of
derivations in a superalgebra will be essential. Such derivations may be 
even/odd depending on whether they preserve parity or not. Even derivations
are defined as usual. By contrast, an odd {\em derivation\/} $D$ of a 
superalgebra satisfies
$$
  D(ab)=\cases{(Da)b+a(Db) & if $a$ is even\cr (Da)b-a(Db) & if $a$ is odd.\cr}
$$
Inner derivations are given by supercommutators $D=[c,\cdot]$ where $c$
is fixed. Moreover, the linear space of all derivations is a Lie superalgebra
since any bracket $[D,D']$ is a derivation, too.

We shall frequently use tensor products.
It is important to realize that tensor products of superalgebras are special.
Generally speaking, if $X$ and $Y$ are $\ZZ_2$-graded algebras, the 
multiplication in $X\otimes Y$ is given by
$$
   (x\otimes y)(x'\otimes y')=
   \cases{-xx'\otimes yy' & if $x'$ and $y$ are odd\cr
            \hphantom{-}xx'\otimes yy' & otherwise.\cr}
$$
In physics, such tensor products are familiar constructions when  
dealing with Fock spaces of different fermions. For, if $E$ and $F$ are
two vector spaces, there is a natural isomorphism
$$    \bw(E\oplus F)\cong\bw E\otimes\bw F\ . $$

\section{Superconnections and the Higgs Field}

Let $M$ now be a (connected, oriented) differentiable manifold. It is 
helpful to think of $M$ as a model of Euclidean spacetime. 
Later, we shall assume that its dimension is even.
By a {\em superbundle} we mean a vector bundle on $M$ whose fibers are 
superspaces. Examples are:
\begin{quote}
            -- the bundle $\bw T^*M$ of exterior differentials,\\
            -- the Clifford bundle $C(M)$ of a Riemannian manifold,\\
            -- the endomorphism bundle of a superbundle.
\end{quote}
Sections of a superbundle $B$ obviously form a superspace $\Gamma(B)$.

The most common object for integration on manifolds is the exterior algebra 
of differential forms (a supercommutative algebra),
$$        \Omega=\Gamma(\bw T^*M).   $$
Elements of $\Omega$ of degree $p$ are said to be $p$-forms on
$M$. They are are even (odd) if $p$ is even (odd). The even elements 
constitute a commutative subalgebra $\Omega^+$ of $\Omega$.
There is a canonical odd derivation $d$ on $\Omega$, commonly known as the 
{\em exterior derivative}, mapping $p$-forms into $(p+1)$-forms such that
$d^2=0$, which reduces to the ordinary derivative $df$ on functions 
$f:M\to\RR$.

In gauge theory one chooses a compact Lie group $G$, called the gauge group,
and some principal $G$ bundle $P$ over $M$ to start from. A vector bundle,
which is an associated $G$ bundle, may be obtained from any representation 
$\rho$ of the group $G$. The choice of $\rho$ is dictated by the multiplet of 
particles (or fields) one wishes to describe. Here we shall be interested in
representations spaces (real or complex) carrying a $\ZZ_2$-grading respected 
by $\rho$. This in particular implies that $\rho$ has subrepresentions 
$\rho^\pm$ of same dimension.

Let $B$ some $G$ superbundle obtained in the above manner. We will then
consider the superspace of $B$-valued differential forms,
$$
   S(B)=\Gamma(\bw T^*M\otimes B),
$$
and also the superalgebra of {\em local operators\/} on $S(B)$,
$$ 
    A(B)=\Gamma(\bw T^*M\otimes{\rm End}\,B).
$$
As opposed to a differential operator, a local operator preserves fibers, that
is to say, it commutes with the multiplication by functions $f\in C^\infty(M)$.
Since the algebra $\Omega$ acts fiberwise on the vector space $S(B)$ in an 
obvious manner, there is a natural embedding $\Omega\to A(B)$. 
The following notion, due to D.\ Quillen, generalizes the concept
of a covariant derivative. See also [4] for details.
\vspace{3mm}\par\noindent
{\em Definition}. A {\em superconnection\/} on $B$ is a (first-order) 
differential operator $\DD$ on $S(B)$ of odd type satisfying the Leibniz rule
$$
          [\DD,\omega)]=d\omega, \qquad \omega\in\Omega\subset A(B).
$$
A few observations are immediate.
\begin{enumerate}
\item If $\DD$ and $\DD'$ are two different superconnections, their difference
supercommutes with $\omega$ and so is a local operator of odd type: 
superconnections form an affine space modelled on the vector space $A^-(B)$.
\item  $\DD^2=\hf[\DD,\DD]$ is even. From the generalized Jacobi identity and 
the relation $[\DD[\DD,\omega]=d^2\omega=0$ we see that $\DD^2$ 
commutes with $i(\omega)$ and hence is a local operator. We call 
$\FF =\DD^2\in A^+(B)$ the {\em curvature\/} of the superbundle $B$.
\item Bianchi's identity $[\DD,\FF] =0$ holds.
\item Any superconnection gives rise to an odd derivation of the superalgebra
$A(B)$, again denoted $\DD$, in a way consistent with the Leibniz rule:
$\DD a =[\DD,a]$ $(a\in A(B))$. Thus, $\DD\FF=0$ is another way to
write Bianchi's identity.
\end{enumerate}
It is not difficult to prove the following structure theorem. Any
superconnection decomposes as $\DD = D+L$ where $D$ is a covariant derivative
on $B$ while $L\in A^-(B)$ (with no further restriction on $L$). Thus, 
$D$ maps $p$-forms into $(p+1)$-forms and, in local coordinates,
$$
        D = dx^\mu\big(\partial_\mu + A_\mu(x)\big)
$$
where $A_\mu(x)$ is the gauge potential, taking values in some representation
of the Lie algebra of $G$, and
\begin{equation}
  L=L(x) + \sum_{p\ge 2}dx^{\mu_1}\wedge\ldots\wedge dx^{\mu_p}
       L_{\mu_1\cdots\mu_p}(x)    \label{L}
\end{equation}
with scalar field $L(x)$ (the $p=0$ contribution) and tensor fields 
$L_{\mu_1\cdots\mu_p}(x)$ of degree $p\ge 2$. Fields in $L$ are thought 
of as sections of the endomorphismen bundle ${\rm End}^-B$ if $p$ is even 
or ${\rm End}^+B$ if $p$ is odd. The idea of superconnections has thus 
provided new fields other than the gauge potential with a definite behavior 
under gauge transformations. We shall refer to the scalar field $L(x)$ as the 
{\em Higgs field\/} of the superconnection $\DD$. At present, we need not
introduce tensor fields of degree $p\ge 2$ in a superconnection if we merely
wish to accommodate the particles of the Standard Model, and we will assume 
from now on that the series (\ref{L}) truncates after the zeroth order term:
$$   L=L(x)\in\Gamma({\rm End}^-B).  $$
With respect to the grading $B=B^+\oplus B^-$, we may
conveniently represent any superconnection as a matrix of operators:
$$
           \DD = \pmatrix{ D^+ & i\Phi^*\cr i\Phi & D^-}\qquad
            L  = \pmatrix{ 0 & i\Phi^*\cr i\Phi & 0}\ .
$$
Clearly, $D^\pm$ are covariant derivatives on $B^\pm$. We also assume
that $B$ is a Hermitian vector bundle and $\DD$ is skew-selfadjoint in 
the sense that
$$
    (\DD v,w)+ (v,\DD w)= d(v,w), \qquad v,w\in S(B)
$$ 
where $(v,w)\in C^\infty(M)$ denotes the induced scalar product of sections.
At each point $x\in M$, the field $\Phi^*(x)$ is the adjoint 
of the field $\Phi(x)$ and may be looked upon as an $n\times n$ matrix 
if the bundle $B$ has rank $2n$ and some frame has been chosen. With no 
further restrictions on $L$, the Higgs field has $n^2$ independent components.

The curvature decomposes as
          $$ \FF =D^2 +[D,L]+L^2  $$
where the 2-form $F=D^2$ is referred to as the {\em field strength}, 
the 1-form $[D,L]$ is the covariant derivative of the Higgs field, and 
the 0-form $L^2$ determines the {\em Higgs potential}, once a scalar
product $(\FF,\FF)$ has been defined (details in [2]).

\section{Constructing the Standard Model}

Assume now that $P$ is a principal $G$ bundle where the gauge group $G$ is 
either the unitary group $U(n)$ or a subgroup thereof. Since $G$ acts on
$P$ but also on $\CC^n$ (equipped with the standard scalar product), 
we may construct the associated $G$ bundle
$$            V = P\times_G \CC^n $$
having fibers isomorphic to $\CC^n$. Though there is no natural
graded structure on $V$, the exterior algebra $B=\bw V$ is in fact a G 
superbundle of rank $2^n$. The representation $\bw$ of $G$ acting on its fibers
respects parity and has subrepresentations $\bw^\pm$ of equal dimension.
By construction, $V$ is a Hermitian vector bundle and so is $\bw V$.
We will be mainly concerned with the following two cases:
$$
  \begin{tabular}[c]{lll}
   $n=2$ & $G=U(2)$        & electro-weak theory \\
   $n=5$ & $G\subset U(5)$ & Standard Model.
  \end{tabular}
$$
To introduce fermions into the theory we need a few more assumptions. 
Let $M$ now be a Riemannian manifold of dimension $2m$ and $C(M)$ be its 
Clifford bundle (canonically associated with the cotangent bundle $T^*M$).
Its construction formalizes Dirac's notion of an "algebra of $\gamma$ 
matrices connected with spacetime". Let $c:T^*M\to{\rm End}\,S$ be a 
spin$^c$-structure on $M$, i.e., 
$S$ is a complex vector bundle of rank $2^m$ on $M$, called the 
{\em spinor bundle}, and the bundle map $c$ satisfies $c(v)^2+(v,v)=0$
with respect to the scalar product $(\cdot,\cdot)$ on $T^*M$ induced by
the Riemannian structure. It may be shown that $c$ extends to an algebraic
isomorphism $C(M)\to{\rm End}\,S$ and thus gives $S$ the structure of a
Clifford module. The $\gamma$ matrices are locally recovered by setting 
$\gamma^\mu=c(dx^\mu)$. Clifford modules formalizes Dirac's concept of a
"space on which the $\gamma$'s act". The eigenvalues $\pm 1$
of the chirality operator $\gamma_5=i^m\gamma^1\gamma^2\cdots\gamma^{2m}$
give $S$ the structure of a superbundle.

In order to incorporate gauge symmetries we consider the {\em twisted
spinor bundle},
$$    E =\bw V\otimes S. $$ 
Since both $S$ and $\bw V$ are superbundles, so is $E$. In particular,
$$    E^+ = (\Wp V\otimes S^+)\,\oplus\, (\Wm V\otimes S^-). $$
Dirac fields describing leptons and quarks are thought of as components of 
{\em one\/} master field $ \psi\in\Gamma(E^+)$. The restriction to $E^+$
couples the helicity of $S$ to the parity of the exterior algebra.
Note that the master field $\psi$ is capable of describing $2^n$ elementary 
fermion fields. Left- and right-handed fields count as different components.
The fact that fermion fields are Grassmann variables in Euclidean field
theory will not be discussed. Nevertheless, the reader should be aware that
$\bar{\psi}(x)$ and $\psi(x)$ anticommute and, contrary to the situation in
Minkowski field theory, are unrelated.

The fermionic part of the Lagrangian is taken to be $\bar{\psi}i{\cal D}\psi$ 
where $\cal D$ is a generalized Dirac operator. We shall not go into the
details here except to say that $\cal D$ is constructed from the
superconnection $\DD$ in very much the same way as the conventional Dirac
operator $D\kern-0.6em/\kern0.2em$ is constructed from the covariant
derivative $D$. Formally, $\cal D$ is a (first-order) differential operator
on $\Gamma(E)$ of odd type satisfying $[{\cal D},f]=c(df)$ for all
$f\in C^\infty(M)$. Being odd in particular means that a generalized Dirac
operator cannot contain a "mass term". Leptons and quarks must acquire 
their masses by the Higgs mechanism. Our Ansatz for the Lagrangian takes
care of both the Yukawa and the gauge coupling of fermions. 

Let us first turn to the $U(2)$ model describing weak isospin (quantum number
$I$) along with hypercharge (quantum number $Y$). It goes without saying that
$U(1)_Y$ is considered the center of the group $U(2)$. But, as a matter of 
convention, the generator of $U(1)_Y$ is taken here as the {\em negative\/} 
hypercharge. Irreducible representations are characterized by $I$ and $Y$
subject to the restriction  $2I+Y=$even.
After symmetry breaking the residual gauge group will be
$$
    U(1)_Q =\left\{\pmatrix{1&0\cr 0&e^{i\alpha}\cr},\ 0\le\alpha<2\pi\right\}
$$
giving rise to the notion of the electric charge $Q$ as a conserved quantity.
Likewise, the generator of $U(1)_Q$ in any representation is taken as $-Q$.
By construction, the charge then satisfies the relation $Q=I_3+\hf Y$ of 
Gell-Mann-Nishijima.
 
The master field $\psi$ in the $U(2)$ model has four components,
      $$  \psi =(\nu_{eR},e_R,\nu_{eL},e_{eL})\in\Gamma(E^+),$$
associated to three invariant subspaces of $\bw\CC^2$. As indicated,
the components describe the electron (and the accompanying neutrino).
There is another master field for the muon and one for the $\tau$ lepton.
Left- and right-handed fields have different properties under gauge 
transformations:
$$
\begin{tabular}[c]{rllll}
$\nu_{eR}\to\bw^0\CC^2$       &\qquad singlet,&\ $Y= 0$&\ $I=0  $&\ $Q= 0$\\
$e_R     \to\bw^2\CC^2$       &\qquad singlet,&\ $Y=-2$&\ $I=0  $&\ $Q=-1$\\
$\nu_{eL},e_{eL}\to\bw^1\CC^2$&\qquad doublet,&\ $Y=-1$&\ $I=\hf$&\ $Q=0,-1$. 
\end{tabular}
$$
The appearance of a right-handed neutrino field, foreign to most weak 
interaction theories, signalizes that the neutrino is assumed to acquire a 
small mass after symmetry breaking.

The bosonic sector has a spin-one gauge field of four components,
corresponding to the photon, the $Z$, and the $W^\pm$. 
In addition, there are two
Higgs doublets of opposite hypercharge. If the Lagrangian is at most 
quadratic in the curvature $\FF$ and gauge invariant, there are only
very few free parameters left that enter the action functional.

We now turn to another Lie group $G$ with Lie algebra 
\begin{equation}
 \mbox{Lie\,}G \cong{\bf su(3)}\oplus{\bf su(2)}\oplus{\bf u(1)} \label{Lie}
\end{equation}
large enough to enable us to incorporate quark fields and strong 
interactions.
In addition, we require that $G$ be a subgroup of $SU(5)$, i.e., we define
\begin{equation}
    G =\{(u,v)\in \mbox{U(3)}\times\mbox{U(2)}\ |\ \det u\cdot\det v=1\}
\end{equation}
and let the embedding $G\to SU(5)$ be given by
$$
                  (u,v) \mapsto\pmatrix{u &0\cr 0&v\cr}\ .
$$
There are in fact three basic symmetry groups involved in our model. Note
that they are related by the following exact sequence
\begin{equation}
     1\arr{}\mbox{SU(3)}\arr{j}G\arr{s}\mbox{U(2)}\arr{}1     \label{exs}
\end{equation}
where $j(u)=(u,1)$ and $s(u,v)=v$.
Though there is the isomorphism (\ref{Lie}) between Lie algebras, the group 
$G$ cannot be identified with the direct product $SU(3)\times U(2)$. 
It is correct to say that the color group $SU(3)$ of quantum 
chromodynamics is embedded in $G$ as a subgroup. But the gauge group 
$U(2)$ of leptons is recovered only as the quotient $G/SU(3)$. This fact
influences our idea of what the hypercharge $Y$ should be. To see the point
more clearly we consider the exact sequence
\begin{equation}
  \label{eq:der}
  1\arr{}\ZZ_3\arr{j}\tilde{U}(1)_Y\arr{s}U(1)_Y\arr{}1
\end{equation}
obtained from (\ref{exs}) by restricting to the centers. In this way we learn
that the group $\tilde{U}(1)_Y$, a threefold cover of $U(1)_Y$, may also be
looked upon as a one-dimensional closed subgroup of the two-torus: 
$$
  \tilde{U}(1)_Y =\{(e^{i\beta},e^{i\alpha})\ |\ 3\beta+2\alpha=
                                               0 \bmod 2\pi\}. 
$$
As before, $U(1)_Y$ defines the hypercharge. So does $\tilde{U}(1)_Y$ by the
local isomorphism $s$ whose inverse is 
\begin{equation}
  \label{eq:inv}
      s^{-1}(e^{i\alpha})=(e^{-i2\alpha/3},e^{i\alpha}).
\end{equation}
Locally, the group $\tilde{U}(1)_Y$ is represented by a phase factor 
$e^{-i\alpha Y}$ in any unitary irreducible representation 
of $G$. Notice, however, that $Y\in\QQ$ in general.

The vector bundle $V$ is now modelled on the fiber space
$$
          \CC^5=\CC^3\oplus\CC^2
$$
with subspaces $\CC^3$ and $\CC^2$ carrying the {\em fundamental 
representations\/} of the color group SU(3) and the weak-isospin group $SU(2)$
respectively. As explained above, passage to the exterior algebra $\bw\CC^5$ 
is very essential, the fiber space of the superbundle $\bw V$  carrying 
the reducible representation $\bw$ of $G$. From the natural isomorphism 
$\bw(\CC^3\oplus\CC^2)\cong\bw\CC^3\otimes\bw\CC^2$ we obtain $\bw(u,v)=
\bw u\otimes\bw v$  for $(u,v)\in G$ and hence
$$
 \textstyle \bw^r(u,v)=\sum_{p+q=r}\bw^pu\otimes\bw^q v,\qquad r=0,\ldots,5.
$$
Consequently, any fundamental fermion (quark or lepton) must belong to one
of the irreducible representations of $G$,
$$       
   \bw^{p,q}=\bw^p\otimes\bw^q\qquad p=0,1,2,3,\qquad q=0,1,2
$$ 
whose dimension is ${p\choose 3}{q\choose 2}$.
To find its hypercharge we use Eq.\ (\ref{eq:inv}):
$$ 
    e^{-i\alpha Y} = \bw^{p,q}\big(s^{-1}(e^{i\alpha})\big)
                   =\exp(-i2p\alpha/3+iq\alpha)
$$
and thus obtain the fundamental relation
\begin{equation}
  \label{eq:pq}
\fbox{\parbox{40mm}{$$\textstyle Y =\frac{2}{3}p-q$$}}\ .
\end{equation}
Clearly, $Y$ is integer valued if $p=0\bmod 3$ (for leptons) 
and fractional otherwise (for quarks). A similar statement holds for the
electric charge $Q$.

The master field $\psi$ has $2^5=32$ components. Dirac fields that enter $\psi$
are characterized by three different "parities" owing to the $\ZZ_2$-gradings
of $\bw\CC^5$, $\bw\CC^3$, and $\bw\CC^2$. Their interpretation is as follows: 
$$
\begin{tabular}{rlrl}
  $p+q =$even &: right-handed \hspace{20mm} &$p+q = $odd &: left-handed\\
  $p = $even  &: matter                     &$p   = $odd &: antimatter\\
  $q = $even  &: singlets                   &$q   = $odd &: doublets.
\end{tabular}
$$
Hence, there are left-handed and right-handed fields of equal number.
Likewise, there are matter fields and antimatter fields of equal number.
Each doublet is accompanied by two singlets. The following table shows the
details.
$$
\begin{tabular}{|c|rr|rr|}\hline
               & $q=0$     & $q=2$ & \multicolumn{2}{c|}{$q=1$} \\ \hline
   $p=0$ &\vsp $\nu_{eR}$  & $e_R$ & $\nu_{eL}$ & $e_L$\\  \hline
               &\vsp $u_{1R}$ & $d_{1R}$ & $u_{1L}$    & $d_{1L}$\\
   $p=2$ &\vsp $-u_{2R}$ & $-d_{2R}$ & $-u_{2L}$    & $-d_{2L}$\\
               &\vsp $u_{3R}$ & $d_{3R}$ & $u_{3L}$    & $d_{3L}$\\ \hline \hline
               &\vsp $d^c_{1L}$ & $u^c_{1L}$ & $d^c_{1R}$ & $-u^c_{1R}$\\
   $p=1$ &\vsp $d^c_{2L}$ & $u^c_{2L}$ & $d^c_{2R}$ & $-u^c_{2R}$\\
               &\vsp $d^c_{3L}$ & $u^c_{3L}$ & $d^c_{3R}$ & $-u^c_{3R}$\\ \hline
   $p=3$ &\vsp $e^c_L$ & $\nu^c_{eL}$ & $e^c_R$ & $-\nu^c_{eR}$ \\ \hline
\end{tabular}
$$
Quarks fields such as $u$(up) and $d$(down) come in three {\em colors\/}: 
$i=1,2,3$. The upper index $\vphantom{d}^c$ is used to indicate antimatter.
For instance, $d^c$ is the charge conjugate field obtained from the Dirac 
field $d$. Charge conjugation passes from $\bw^{p,q}$ to $\bw^{3-p,2-q}$  
and therefore reverses the electric charge, the hypercharge, and the helicity:
$$     
          d^c_L :=(d^c)_L =(d_R)^c,\qquad d^c_R:=(d^c)_R=(d_L)^c.
$$
Obviously, the operations $C$, $P$ (charge conjugation and parity) are well
defined on $\psi$, though they need not be symmetries. The introduction
of fields together with their charge conjugates in one multiplet is welcome,
because it eliminates $\bar{\psi}$ as an independent variable in the
Lagrangian. Contrary to a wide-spread assumption, the fields for the
antiquarks, in the present scheme, are assigned the defining representation 
{\bf 3} of $SU(3)$,
while the fields for the quarks are assigned the complex conjugate
representation $\bar{\bf 3}$. Interchanging the role of the two
fundamental representations, however, has no physical implication.

We emphasize once more
that there is a natural place for the right-handed neutrino field, $\nu_{eR}$,
as well as for and its charge conjugate field, $\nu^c_{eL}$.
Both will be needed if the neutrino acquires a nonzero mass. 
The $SU(5)$ gauge model of Georgi and Glashow, however, discards this 
possibility leaving the representations $\bw^{0,0}$ and $\bw^{3,2}$ 
(trivial representations of $SU(3)\times SU(2)$) unoccupied. It seems that
nature provides several generations of fundamental fermions. We offer no
explanation for this fact, but mention that each generation has to be 
introduced by a separate master field.

We have presented a systematic and well-motivated analysis of some
structural aspects of the Standard Model, leaving out all quantitative
results: some of them have already been published [2]. Others are deferred to
the future investigations.

\vspace{5mm}\par\noindent
{\bf References}
\begin{enumerate}
\item D.\ Quillen, Topology {\bf 24},89 (1985)\\
      V.\ Mathai, D.\ Quillen, Topology {\bf 25}, 85 (1986);
\item G.\ Roepstorff, Superconnections and the Higgs Field, hep-th/9801040\\
      published in J.Math.Phys.\ {\bf 40}, 2698 (1999) 
\item G.\ Roepstorff, Superconnection and Matter, hep-th/9801045
\item N.\ Berline, E.\ Getzler, M.\ Vergne: Heat Kernels and Dirac Operators,
          Springer, Berlin 1996      
\end{enumerate}
\end{document}